\documentclass[twocolumn,A4]{article}
\usepackage[dvips]{graphics}
\begin{document}
\onecolumn
\begin{center}
{\bf{\Large Persistent current in a mesoscopic cylinder: effects of 
radial magnetic field}}\\
~\\
Santanu K. Maiti$^{1,2,*}$ and F. Aeenehvand$^3$ \\
~\\
{\em $^1$Theoretical Condensed Matter Physics Division,
Saha Institute of Nuclear Physics, \\
1/AF, Bidhannagar, Kolkata-700 064, India \\
$^2$Department of Physics, Narasinha Dutt College,
129, Belilious Road, Howrah-711 101, India \\
$^3$Faculty of Science, Department of Physics,
Islamic Azad University, Karaj Branch, Iran} \\
~\\
{\bf Abstract}
\end{center}
In this work, we study persistent current in a mesoscopic cylinder
subjected to both longitudinal and transverse magnetic fluxes. A simple 
tight-binding model is used to describe the system, where all the 
calculations are performed exactly within the non-interacting electron 
picture. The current $I$ is investigated numerically concerning its 
dependence on total number of electrons $N_e$, system size $N$, 
longitudinal magnetic flux $\phi_l$ and transverse magnetic flux 
$\phi_t$. Quite interestingly we observe that typical current amplitude
oscillates as a function of the transverse magnetic flux, associated
with the energy-flux characteristics, showing $N \phi_0$ flux-quantum
periodicity, where $N$ and $\phi_0$ $(=ch/e)$ correspond to the system
size and the elementary flux-quantum respectively. This analysis may 
provide a new aspect of persistent current for multi-channel cylindrical
systems in the presence of radial magnetic field $B_r$, associated 
with the flux $\phi_t$.
\vskip 1cm
\begin{flushleft}
{\bf PACS No.}: 73.23.-b; 73.23.Ra; 75.20.-g  \\
~\\
{\bf Keywords}: Mesoscopic cylinder; Persistent current; Radial
magnetic field.
\end{flushleft}
\vskip 4in
\noindent
{\bf ~$^*$Corresponding Author}: Santanu K. Maiti

Electronic mail: santanu.maiti@saha.ac.in
\newpage
\twocolumn

\section{Introduction}

In thermodynamic equilibrium, a small metallic ring threaded by a magnetic 
flux $\phi$ supports a current that does not decay dissipatively even at 
non-zero temperature. It is the so-called persistent current in mesoscopic 
normal metal rings. This phenomenon is a purely quantum mechanical effect, 
and provides an exact demonstration of the Aharonov-Bohm$^1$ effect. In the 
very early days of quantum mechanics, Hund$^2$ predicted the appearance of 
persistent current in a normal metal ring, but the experimental evidences 
of it came much later only after realization of the mesoscopic systems. 
In $1983$, B\"{uttiker} {\em et al.}$^3$ showed theoretically that 
persistent current can exist in mesoscopic normal metal rings threaded by 
a magnetic flux even in the presence of disorder. Few years later, Levy 
{\em et al.}$^4$ first performed the excellent experiment and gave the 
evidence of persistent current in the mesoscopic normal metal rings. 
Following with this, the existence of persistent current was further
confirmed by many experiments.$^{5-9}$ Though there exists a vast literature 
of theoretical$^{10-25}$ as well as experimental$^{4-9}$ results on persistent 
currents, but lot of controversies are still present between the theory and 
experiment. For our illustrations, here we mention some them as follow.
(i) The main controversy appears in the determination of the current
amplitude. It has been observed that the measured current amplitude 
exceeds an order of magnitude than the theoretical estimates. Many efforts 
have been paid to solve this problem, but no
such proper explanation has been found out. Since normal metals are 
intrinsically disordered, it was believed that electron-electron 
correlation can enhance the current amplitude by homogenize the system. 
But the inclusion of the electron correlation$^{26}$ doesn't give any 
significant enhancement of the persistent current. Later, in some recent
papers$^{27-29}$ it has been studied that the simplest nearest-neighbor 
tight-binding model with electron-electron interaction cannot explain the 
actual mechanisms. The higher order hopping integrals in addition to the 
nearest-neighbor hopping integral have an important role to magnify the 
current amplitude in a considerable amount. With this prediction some
discrepancies can be removed, but the complete mechanisms are yet to be
understood. (ii) The appearance of different flux-quantum periodicities 
rather than simple $\phi_0$ ($\phi_0=ch/e$, the elementary flux-quantum) 
periodicity in persistent current is not quite clear to us. The presence 
of other flux-quantum periodicities has already been reported in many 
papers,$^{30-33}$ but still there exist so many conflict. (iii) The 
prediction of the sign of low-field currents is a major challenge in
this area. Only for a single-channel ring, the sign of the low-field 
currents can be mentioned exactly.$^{33-34}$ While, in all other cases
i.e., for multi-channel rings and cylinders, the sign of the low-field
currents cannot be predicted exactly. It then depends on the total number
of electrons ($N_e$), chemical potential ($\mu$), disordered configurations,
etc. Beside these, there are several other controversies those are 
unsolved even today. 

In the present paper, we will investigate the behavior of persistent 
currents in a thin cylinder (see Fig.~\ref{strip}) in the presence of both 
longitudinal and transverse magnetic fluxes, $\phi_l$ and $\phi_t$ 
respectively. Our numerical study shows that typical current amplitude 
in the cylinder oscillates as a function of the transverse magnetic flux 
$\phi_t$, associated with the magnetic field $B_r$, showing $N\phi_0$ 
flux-quantum periodicity instead of simple $\phi_0$-periodicity, where 
$N$ corresponds to the size of the cylinder. This oscillatory behavior 
provides an important signature in this particular study. To the best of 
our knowledge, this phenomenon of periodicity in persistent current has 
not been addressed earlier in the literature.

We organize the paper as follow. In Section $2$, we present the model 
and the theoretical formulations for our calculations. Section $3$ 
discusses the significant results, and finally we summarize our results 
in Section $4$.

\section{Model and the theoretical description}

\begin{figure}[ht]
{\centering \resizebox*{7.5cm}{6cm}{\includegraphics{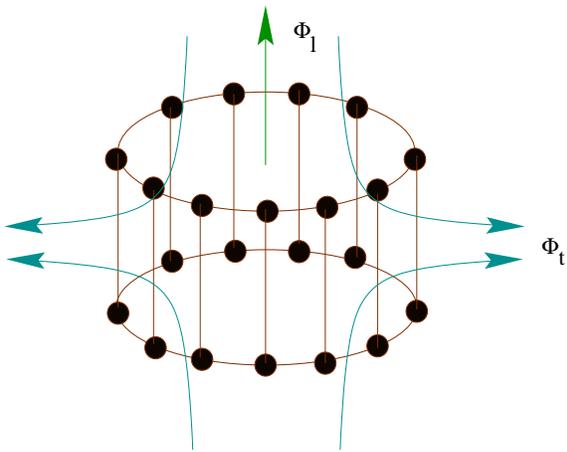}}\par}
\caption{Schematic view of a mesoscopic cylinder subjected to both the 
longitudinal and transverse magnetic fluxes $\phi_l$ and $\phi_t$ 
respectively. Filled circles correspond to the position of the atomic 
sites (for color illustration, see the web version).}
\label{strip}
\end{figure}

Let us refer to Fig.~\ref{strip}. A thin metallic cylinder is subjected to
the longitudinal magnetic flux $\phi_l$ and to the transverse magnetic
flux $\phi_t$. For our illustration, we consider this simplest cylinder,
where only two isolated one-channel rings are connected by some vertical
bonds. The transverse magnetic flux $\phi_t$ is expressed in terms of 
the radial magnetic field $B_r$ by the relation $\phi_t=B_r L d$,
where the symbols $L$ and $d$ correspond to the circumference of each ring 
and the hight of the cylinder respectively. The system of our concern can 
be modeled by a single-band tight-binding Hamiltonian, and in the 
non-interacting picture, it looks in the form,
\begin{eqnarray}
H & = & \sum_{i=1}^N \epsilon_i^L c_i^{L\dagger} c_i^L  
+ \sum_{i=1}^N \epsilon_i^U c_i^{U\dagger} c_i^U \nonumber \\
  & + & v_l^L \sum_{<ij>}
\left[e^{i\left( \theta_1 - \theta_2 \right)} c_i^{L\dagger} c_j^L
+ e^{-i \left( \theta_1 - \theta_2 \right)} c_j^{L\dagger} 
c_i^L \right] \nonumber \\
  & + & v_l^{U} \sum_{<ij>}
\left[e^{i\left( \theta_1 + \theta_2 \right)} c_i^{U\dagger} c_j^U
+ e^{-i \left( \theta_1 + \theta_2 \right)} c_j^{U\dagger} 
c_i^U \right] \nonumber \\
 & + & v_t \sum_{i=1}^N \left(c_i^{L\dagger}c_i^U + c_i^{U\dagger}
c_i^L \right)
\label{hamil}
\end{eqnarray}
In the above Hamiltonian ($H$), $\epsilon_i^L$'s ($\epsilon_i^U$'s) are
the site energies in the lower (upper) ring, $c_i^{L\dagger}$ 
($c_i^{U\dagger}$) is the creation operator of an electron at site $i$
in the lower (upper) ring, and the corresponding annihilation operator for 
this site $i$ is denoted by $c_i^L$ ($c_i^U$). The symbol $v_l^L$ ($v_l^U$)
gives the nearest-neighbor hopping integral in the lower (upper) ring,
while the parameter $v_t$ corresponds to the transverse hopping strength
between the two rings of the cylinder. $\theta_1$ and $\theta_2$ are the
two phase factors those are related to the longitudinal and transverse 
fluxes by the expressions, $\theta_1=2 \pi \phi_l/N \phi_0$ and 
$\theta_2=\pi \phi_t/N \phi_0$, where $N$ represents the total number of
atomic sites in each ring.

At absolute zero temperature ($T=0$ K), the longitudinal persistent 
current so-called the Aharonov-Bohm persistent current in the cylinder 
can be expressed as,
\begin{equation}
I(\phi_l) = - \frac{\partial{E_0(\phi_l,\phi_t})}{\partial{\phi_l}}
\label{current}
\end{equation}
where, $E_0(\phi_l,\phi_t)$ represents the ground state energy. 
We evaluate this 
energy exactly to understand unambiguously the anomalous behavior of 
persistent current, and this is achieved by exact diagonalization of the 
tight-binding Hamiltonian Eq.~(\ref{hamil}). Throughout the calculations, 
we take the site energies $\epsilon_i^L=\epsilon_i^U=0$, which reveal a 
perfect cylinder, the hopping integrals $v_l^L=v_l^U=v_t=2.5$, and for 
simplicity, we use the units where $c=1$, $e=1$ and $h=1$.

\section{Results and discussion}

\begin{figure}[ht]
{\centering \resizebox*{7.75cm}{5cm}{\includegraphics{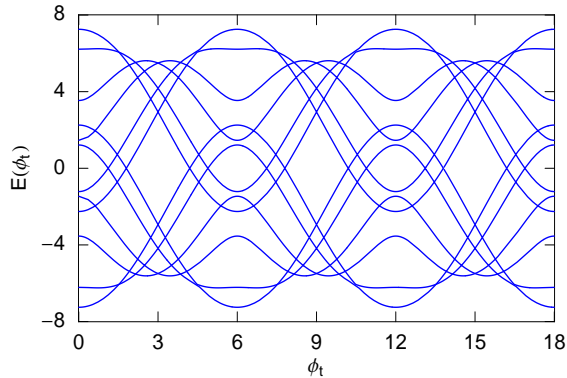}}\par}
\caption{Variation of the energy levels as a function of the transverse
magnetic flux $\phi_t$ for a mesoscopic cylinder with $N=6$. The longitudinal 
magnetic flux $\phi_l$ is set to $0.3$ (for color illustration, see the web 
version).}
\label{traneng}
\end{figure}

To reveal the basic mechanisms of the transverse magnetic flux $\phi_t$ 
on the persistent current, here we present all the results only for the 
non-interacting electron picture. With this assumption, the model becomes 
quite simple and all the basic features can be well understood.
Another realistic assumption is that, we focus on the perfect cylinders
only i.e., the site energies are taken as $\epsilon_i^L=\epsilon_i^U=0$
for all $i$.
\begin{figure}[ht]
{\centering \resizebox*{7.75cm}{5cm}{\includegraphics{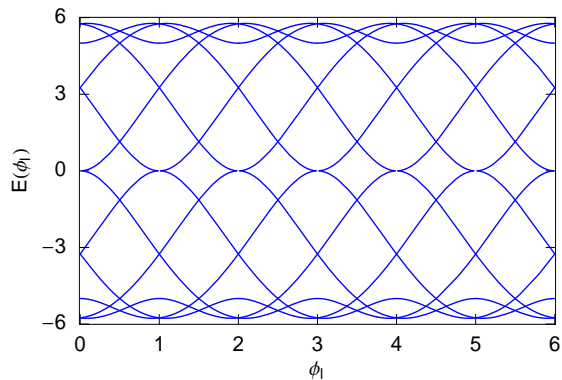}}\par}
\caption{Variation of the energy levels as a function of the longitudinal
magnetic flux $\phi_l$ for a mesoscopic cylinder with $N=6$. The transverse 
magnetic flux $\phi_t$ is set to $4$ (for color illustration, see the web 
version).}
\label{longeng}
\end{figure}
To illustrate the behavior of the persistent current in our concerned system, 
let us first explain the energy-flux characteristics of a mesoscopic cylinder
subjected to both $\phi_l$ and $\phi_t$. Figure~\ref{longeng} 
illustrates the variation of the energy levels as a function of $\phi_l$ 
for such a mesoscopic cylinder with $N=6$.
In this case, the transverse magnetic flux $\phi_t$ is set to $4$. The 
spectrum shows that the energy levels have extrema i.e., either a maxima
or a minima at half-integer or integer multiples of the elementary 
flux-quantum $\phi_0$ ($=1$, for our chosen units). At these extrema 
points the persistent current vanishes since it is evaluated from the first
derivative of the energy eigenstate with respect to the flux $\phi_l$
(Eq.~(\ref{current})). All these energy levels vary periodically with 
$\phi_l$, showing $\phi_0$ flux-quantum periodicity, as expected. This 
$\phi_0$-periodicity cannot be clearly understood from the spectrum 
(Fig.~\ref{longeng}) since the individual energy levels overlap with each 
other and form a complicated picture. The $\phi_l$-dependence of the 
energy levels and their periodicity are quite familiar to us. The
significant behavior appears only when we plot the variation of the 
energy levels as a function of the transverse magnetic flux $\phi_t$. 
In Fig.~\ref{traneng}, we display the dependence of the energy levels 
with $\phi_t$ for a typical mesoscopic cylinder with $N=6$. The 
longitudinal magnetic flux $\phi_l$ is fixed at $0.3$. All the energy 
levels get modified enormously with $\phi_t$, and the dependence of them 
also changes quite a significant way compared to the energy levels plotted 
in the previous spectrum (Fig.~\ref{longeng}). The locations of the 
extrema points of these energy levels no longer situate at the same points 
as obtained in Fig.~\ref{longeng}. In this spectrum (Fig.~\ref{traneng}), 
all the energy levels vary periodically with $\phi_t$, and most interestingly 
we observe that the energy levels exhibit $N\phi_0$ flux-quantum periodicity, 
instead of $\phi_0$. Since in this particular cylinder we choose $N=6$ for
each ring, the energy levels provide $6\phi_0$ ($=6$) flux-quantum 
periodicity with $\phi_t$. The signature of this $N\phi_0$ periodicity 
becomes much more clearly visible from our study of the $I_{typ}$ vs. 
$\phi_t$ characteristics, which we shall describe at the end of this section.

\begin{figure}[ht]
{\centering \resizebox*{7.75cm}{10cm}{\includegraphics{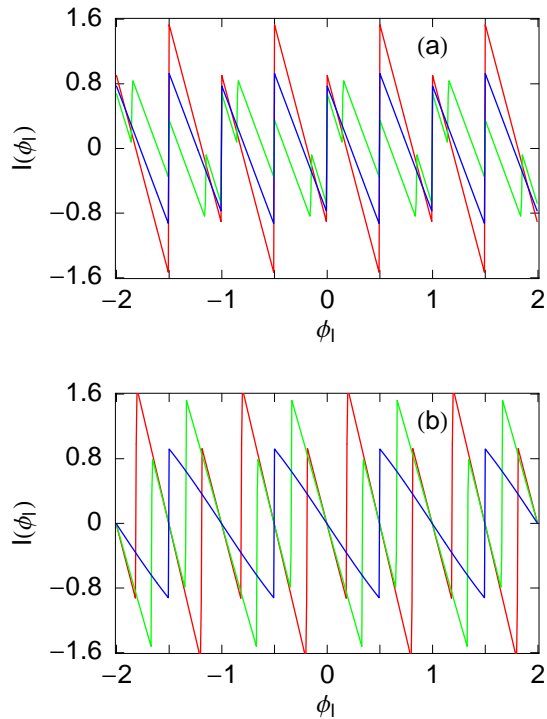}}\par}
\caption{Persistent currents as a function of the longitudinal magnetic 
flux $\phi_l$ for a mesoscopic cylinder with $N=20$. The currents are 
computed for the fixed number of electrons $N_e$, where (a) $N_e=15$ and 
(b) $N_e=18$. The red, green and blue curves correspond to $\phi_t=0$, $4$ 
and $10$, respectively (for color illustration, see the web version).}
\label{longcurr}
\end{figure}

Following the above discussion, next we concentrate our study on the 
current-flux characteristics and the dependence of the current on the 
transverse magnetic flux $\phi_t$. We evaluate all the currents only for
those cylinders which contain fixed number of electrons $N_e$. The current 
carried by an energy eigenstate is obtained by taking the first order 
derivative of the energy for that particular state with respect to the 
flux $\phi_l$, and thus, for the $n$-th energy eigenstate of energy 
$E_n$ (say) the current can be expressed by the relation, 
$I_n(\phi_l)=\partial E_n/\partial \phi_l$. At absolute zero temperature
($T=0$ K), the total persistent current becomes the sum of the 
individual contributions from the lowest $N_e$ energy eigenstates. The 
behavior of the current-flux characteristics for a impurity free mesoscopic
cylinder with $N=20$ is shown in Fig.~\ref{longcurr}, where (a) and (b)
correspond to the currents for $N_e=15$ (odd $N_e$) and $18$ (even $N_e$)
respectively. The red, green and blue curves represent the results for
$\phi_t=0$, $4$ and $10$ respectively. The current exhibits a saw-tooth 
like nature with sharp transitions at several points of $\phi_l$. This 
is due to the existence of the degenerate energy eigenvalues at these 
respective flux points. Depending on the choices of the total number of
electrons $N_e$, the kink appears at different values of $\phi_l$, as 
expected for a multi-channel system.$^{34}$ All these kinks disappear as 
long as we introduce impurities in the system. This phenomenon is very 
well established in the literature and due the obvious reason we do not 
describe further the effect of impurities on the persistent current in 
the present manuscript. From a careful investigation it is observed that 
the current amplitude for a typical value of $\phi_l$ can be controlled 
very nicely by tuning the transverse magnetic flux $\phi_t$.
\begin{figure}[ht]
{\centering \resizebox*{7.75cm}{14cm}{\includegraphics{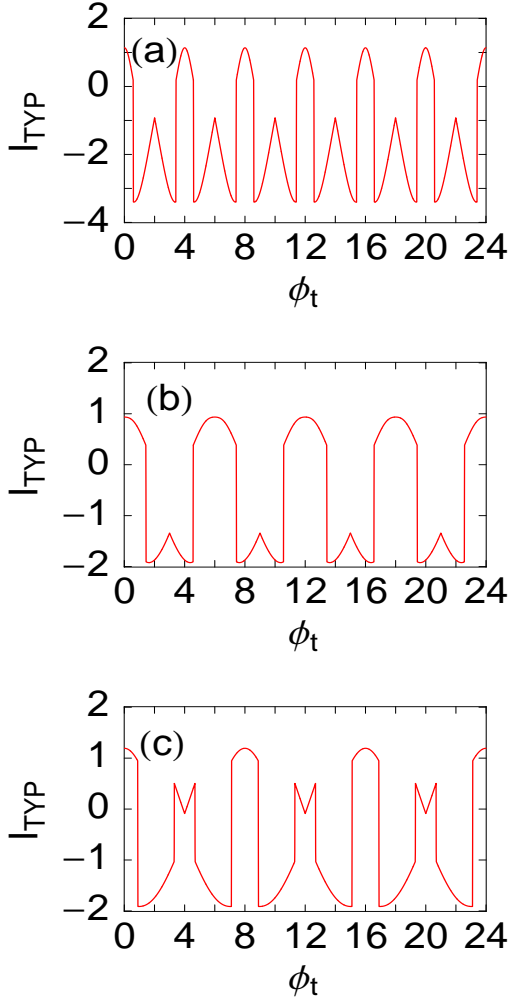}}\par}
\caption{Typical current amplitudes as a function of the transverse magnetic 
flux $\phi_t$ for the three different mesoscopic cylinders, where (a) $N=4$, 
(b) $N=6$ and (c) $N=8$, respectively. The parameter $\phi_l$ is set to 
$0.25$ (for color illustration, see the web version).}
\label{trancurr}
\end{figure}
Now to have a deeper insight to the effects of the transverse magnetic flux 
on persistent current, we focus our study on the variation of the typical
current amplitude ($I_{typ}$) with the magnetic flux $\phi_t$. As
illustrative example, in Fig.~\ref{trancurr} we display the variation of 
the typical current amplitude $I_{typ}$ as a function of $\phi_t$, where
the parameter $\phi_l$ is fixed to $0.25$. Figures~\ref{trancurr}(a), (b)
and (c) correspond to the results for the cylinders with $N=4$, $6$ and $8$
respectively. Quite interestingly we see that the typical current amplitude
varies periodically with $\phi_t$ providing $N\phi_0$ flux-quantum 
periodicity, instead of the conventional $\phi_0$ periodicity. Thus for the 
cylinder with $N=4$, the current exhibits $4\phi_0$ ($=4$) periodicity, while 
it becomes $6\phi_0$ ($=6$) for the cylinder with $N=6$ and $8\phi_0$ ($=8$)
for the cylinder with $N=8$. From these results we clearly observe that the 
variation of $I_{typ}$ with $\phi_t$ within a period for the cylinder with 
$N=8$ is exactly similar to that of the cylinders with $N=6$ and $N=4$ 
within their single periods. Such an $N\phi_0$ periodicity is just the 
replica of the $E$ versus $\phi_t$ characteristics which we have described 
earlier. This phenomenon is really a very interesting one and may provide 
a new aspect of persistent current for multi-channel cylinders in the 
presence of the radial magnetic field $B_r$.

\section{Concluding remarks}

In conclusion, we have studied persistent currents in a mesoscopic 
cylinder subjected to the longitudinal magnetic flux $\phi_l$ and to the
transverse magnetic flux $\phi_t$. We have used a simple tight-binding model
to describe the system and calculated all the results exactly within the 
non-interacting electron picture. Quite interestingly we have observed that 
the typical current amplitude oscillates as a function of the transverse 
magnetic flux $\phi_t$, associated with the energy-flux characteristics, 
providing $N\phi_0$ flux-quantum periodicity. This phenomenon is completely 
different from the conventional oscillatory behavior, like as we have got 
for the case of $I$ versus $\phi_l$ characteristic which shows simple 
$\phi_0$ flux-quantum periodicity. This study may provide a new aspect of 
persistent current for multi-channel cylindrical systems in the presence 
of the radial magnetic field $B_r$.

This is our first step to describe how the persistent current in a thin
cylinder can be controlled very nicely by means of the transverse magnetic 
flux. Here we have made several realistic assumptions by ignoring the 
effects of the electron-electron correlation, disorder, temperature, 
chemical potential, etc. All these effects can be incorporated quite 
easily with this present formalism and we need further study in such 
systems.

\vskip 0.3in
\noindent
{\bf\Large Acknowledgments}
\vskip 0.2in
\noindent
I acknowledge with deep sense of gratitude the illuminating comments and
suggestions I have received from Prof. Arunava Chakrabarti and Prof.
S. N. Karmakar during the calculations.


\begin{thebibliography}{99}

\bibitem{ab} Y. Aharonov and D. Bohm, Phys. Rev. 115, 485 (\textbf{1959}).
\bibitem{hund} F. Hund, Ann. Phys. (Leipzig) 32, 102 (\textbf{1938}).
\bibitem{butt} M. B\"{u}ttiker, Y. Imry, and R. Landauer, Phys. Lett. A
96, 365 (\textbf{1983}).
\bibitem{levy} L. P. Levy, G. Dolan, J. Dunsmuir, and H Bouchiat, Phys. Rev.
Lett. 64, 2074 (\textbf{1990}).
\bibitem{mailly1} D. Mailly, C. Chapelier, and A. Benoit, Phys. Rev. Lett.
70, 2020 (\textbf{1993}).
\bibitem{chand} V. Chandrasekhar, R. A. Webb, M. J. Brady, M. B. Ketchen,
W. J. Gallagher, and A. Kleinsasser, Phys. Rev. Lett. 67, 3578 
(\textbf{1991}).
\bibitem{jari} E. M. Q. Jariwala, P. Mohanty, M. B. Ketchen, and R. A. Webb,
Phys. Rev. Lett. 86, 1594 (\textbf{2001}).
\bibitem{yu} N. Yu and M. Fowler, Phys. Rev. B 45, 11795 (\textbf{1992}).
\bibitem{deb} R. Deblock, R. Bel, B. Reulet, H. Bouchiat, and D. Mailly,
Phys. Rev. Lett. 89, 206803 (\textbf{2002}).
\bibitem{butt1} M. B\"{u}ttiker, Phys. Rev. B 32, 1846 (\textbf{1985}).
\bibitem{cheu1} H-F Cheung, E. K. Riedel, and Y. Gefen, Phys. Rev. Lett.
62, 587 (\textbf{1989}).
\bibitem{cheu2} H. F. Cheung, Y. Gefen, E. K. Riedel, and W. H. Shih,
Phys. Rev. B 37, 6050 (\textbf{1988}).
\bibitem{land} R. Landauer and M. B\"{u}ttiker, Phys. Rev. Lett. 54,
2049 (\textbf{1985}).
\bibitem{byers} N. Byers and C. N. Yang, Phys. Rev. Lett. 7,
46 (\textbf{1961}).
\bibitem{von} F. von Oppen and E. K. Riedel, Phys. Rev. Lett. 66,
84 (\textbf{1991}).
\bibitem{mont} G. Montambaux, H. Bouchiat, D. Sigeti, and R. Friesner,
Phys. Rev. B 42, 7647 (\textbf{1990}).
\bibitem{bouc} H. Bouchiat and G. Montambaux, J. Phys. (Paris)
50, 2695 (\textbf{1989}).
\bibitem{alts} B. L. Altshuler, Y. Gefen, and Y. Imry, Phys. Rev. Lett.
66, 88 (\textbf{1991}).
\bibitem{schm} A. Schmid, Phys. Rev. Lett. 66, 80 (\textbf{1991}).
\bibitem{san11} S. K. Maiti, Int. J. Mod. Phys. B 22, 4951 (\textbf{2008}).
\bibitem{abra} M. Abraham and R. Berkovits, Phys. Rev. Lett. 70,
1509 (\textbf{1993}).
\bibitem{mull} A. M\"{u}ller-Groeling and H. A. Weidenmuller, Phys. Rev. B
49, 4752 (\textbf{1994}).
\bibitem{kulik1} I. O. Kulik, Physica B 284, 1880 (\textbf{2000}).
\bibitem{orella1} P. A. Orellana, M. L. Ladron de Guevara, M. Pacheco,
and A. Latge, Phys. Rev. B 68, 195321 (\textbf{2003}).
\bibitem{kulik2} I. O. Kulik, JETP Lett. 11, 275 (\textbf{1970}).
\bibitem{san3} S. K. Maiti, J. Chowdhury, and S. N. Karmakar, Solid State
Commun. 135, 278 (\textbf{2005}).
\bibitem{san4} S. K. Maiti, J. Chowdhury, and S. N. Karmakar, Synthetic
Metals 155, 430 (\textbf{2005}).
\bibitem{san10} S. K. Maiti, Int. J. Mod. Phys. B 21, 179 (\textbf{2007}).
\bibitem{san5} S. K. Maiti, J. Chowdhury, and S. N. Karmakar, J. Phys.:
Condens Matter 18, 5349 (\textbf{2006}).
\bibitem{avishai} K. Yakubo, Y. Avishai, and D. Cohen, Phys. Rev. B
67, 125319 (\textbf{2003}).
\bibitem{weiden} E. H. M. Ferreira, M. C. Nemes, M. D. Sampaio, and
H. A. Weidenm\"{u}ller, Phys. Lett. A 333, 146 (\textbf{2004}).
\bibitem{san8} S. K. Maiti, Int. J. Mod. Phys. B 21, 3001 (\textbf{2007});
[Addendum: Int. J. Mod. Phys. B 22, 2197 (\textbf{2008})].
\bibitem{san7} S. K. Maiti, Phy. Scr. 73, 519 (\textbf{2006});
[Addendum: Phy. Scr. 78, 019801 (\textbf{2008})].
\bibitem{san9} S. K. Maiti, Physica E 31, 117 (\textbf{2006}).

\end{thebibliography}
\end{document}